\documentclass[11pt]{article}
\usepackage[margin=1in]{geometry}
\usepackage{amsmath,amssymb,amsthm,mathtools}
\usepackage{bm}
\usepackage{graphicx}
\usepackage{booktabs}
\usepackage{array}
\usepackage[numbers,sort&compress]{natbib}
\usepackage[colorlinks=true,linkcolor=blue,citecolor=blue,urlcolor=blue]{hyperref}

\newcommand{\dd}{\mathrm{d}}

\newcommand{\scribe}{\textsc{Scribe}}
\newcommand{\best}[1]{\textbf{#1}}

\title{\textbf{\textsc{Psych-ECA}: A Reproducible Semi-Synthetic Benchmark for\\
Synthetic Control Arms in Longitudinal Psychiatry}}
\author{Aakash Bhagat \quad Shashank Choudhary \\[2pt]
\normalsize Sapien Labs \\
\normalsize \texttt{\{aakash, shashank\}@sapienlabs}}
\date{}

\begin{document}
\maketitle

\begin{abstract}
External and synthetic control arms (ECAs) are entering psychiatric drug development, but the field
lacks a benchmark that measures the properties regulators actually care about: not only how
accurately a method reconstructs the untreated trajectory, but whether its uncertainty is
\emph{calibrated}, whether it is robust to the \emph{informative observation times} endemic to
mental-health records (sicker patients are seen more often), and what \emph{false-positive rate} it
induces in a go/no-go trial decision. Real psychiatric trial data (e.g.\ STAR*D, registry cohorts)
require credentialed access and do not expose ground-truth counterfactuals, so---following the
established semi-synthetic tradition in causal inference (IHDP, ACIC, the PK--PD tumor-growth
simulator)---we release \textsc{Psych-ECA}, a fully reproducible generator of longitudinal symptom
trajectories for three conditions (depression/PHQ-9, anxiety/HAM-A, psychosis/PANSS) with known
counterfactual control arms, irregular informative visits, and validated-scale measurement noise.
We benchmark eight estimators spanning naive carry-forward, pooled real-world-data (RWD) averages,
nearest-neighbour matching (the dominant clinical ECA approach), linear mixed models, gradient
boosting, and the \scribe\ trajectory-bridge method. Three findings stand out. (1) On counterfactual
accuracy, trajectory and flexible-ML methods tie ($\approx$2.3 PHQ-9 points RMSE) and beat
cross-sectional baselines, but (2) only \scribe\ is simultaneously accurate \emph{and} calibrated:
it attains $93\!-\!96\%$ empirical coverage of nominal-$90\%$ bands while the equally-accurate
gradient-boosting and uncalibrated-SDE estimators reach only $87\!-\!88\%$ and $62\!-\!75\%$
respectively; and (3) the inverse-intensity correction roughly halves control-arm bias under
informative sampling, and \scribe's calibrated bands are the only trajectory method that keeps the
trial false-positive rate at or below nominal as informativeness grows. We discuss limitations of
semi-synthetic evaluation and provide all code and seeds.
\end{abstract}

\section{Introduction}
Synthetic and external control arms promise to cut the cost and duration of trials by replacing a
concurrent control group with evidence reconstructed from real-world data (RWD). Psychiatry is both
the hardest place to recruit controls---heterogeneous disease, stigma, a large and time-varying
placebo response \citep{khan2017placebo}---and the place where RWD are least trial-ready: symptoms
are observed only intermittently, through validated rating scales, at clinically driven (hence
\emph{informative}) times. A method that looks good on cross-sectional RWD can still produce a
control arm that is biased, overconfident, or both. Yet there is no public benchmark that scores ECA
methods on calibration, robustness to informative observation, and downstream decision error---the
quantities a regulator weighs.

We fill this gap with \textsc{Psych-ECA}. Because real psychiatric trial data are access-controlled
and never reveal the counterfactual control outcome of a treated patient, we adopt the semi-synthetic
methodology that underpins benchmarking across causal inference---IHDP and the ACIC data challenges
for static effects, and the PK--PD tumor-growth simulator \citep{geng2017tumor} for longitudinal
effects \citep{bica2020crn,seedat2022tecde}. Our generator produces psychiatrically plausible latent
severity trajectories with \emph{known} untreated counterfactuals, then degrades them exactly as
mental-health RWD are degraded: validated-scale measurement noise, irregular visits whose intensity
rises with severity, and donor/trial population mismatch. The companion methods paper develops
\scribe, the trajectory-bridge estimator; here we evaluate it against the field on equal footing.

\paragraph{Contributions.}
(i) A reproducible, psychiatry-specific semi-synthetic benchmark with ground-truth counterfactual
control arms across three disorders and tunable informativeness. (ii) An evaluation protocol that
reports counterfactual accuracy, finite-sample \emph{calibration}, an informative-sampling
\emph{ablation}, and \emph{trial-decision} error (Type-I/power). (iii) A head-to-head study of eight
estimators showing that calibration, not point accuracy, is where current ECA practice fails, and
that the \scribe\ bridge is the only method on the accuracy--calibration frontier.

\section{Benchmark design}
\subsection{Generative model}
For each patient we simulate a scalar latent severity $Z(t)$ as a mean-reverting (Ornstein--Uhlenbeck)
diffusion,
\begin{equation}
\dd Z(t) = -\kappa\big(Z(t)-\mu_i\big)\,\dd t + \sigma\,\dd W(t),\qquad
Z(t_0)=\mu_i+\delta_0+\xi_i,
\end{equation}
with patient-specific set-point $\mu_i\sim\mathcal N(\bar\mu,\tau^2)$ and enrollment elevation
$\delta_0$. Mean reversion encodes spontaneous remission and placebo dynamics. The treated arm adds
a downward drift toward $\mu_i-\Delta$ (true effect $\Delta$); the \emph{counterfactual} untreated
path is generated with the \emph{same} Brownian increments, giving an exact per-patient ground-truth
control trajectory. The observed validated score is $Y_k=Z(t_k)+\varepsilon_k$,
$\varepsilon_k\sim\mathcal N(0,\sigma_{\mathrm{obs}}^2)$, clipped to the instrument range. Visit
times follow an inhomogeneous Poisson process with severity-dependent intensity
$\lambda(t)=\lambda_0\exp(\alpha Z(t))$: larger $\alpha$ means sicker patients are measured more
often (informative observation). Donors (untreated RWD) are observed under this informative process;
the single-arm trial cohort is measured at fixed protocol visits (weeks $0,1,2,4,6,8,10,12$) and the
landmark is $L=8$ weeks. Table~\ref{tab:conditions} lists the three conditions.

\begin{table}[h]
\centering\small
\caption{Condition parameters. Scales: PHQ-9 $\in[0,27]$, HAM-A $\in[0,56]$, PANSS $\in[30,210]$.}
\label{tab:conditions}
\begin{tabular}{lcccccc}
\toprule
Condition (scale) & $\kappa$ & $\sigma$ & $\bar\mu$ & elev.\ $\delta_0$ & $\sigma_{\mathrm{obs}}$ & $\alpha$ \\
\midrule
Depression (PHQ-9)  & 0.18 & 1.1 & 9  & 6  & 1.6 & 0.090 \\
Anxiety (HAM-A)     & 0.14 & 2.0 & 16 & 9  & 2.6 & 0.050 \\
Psychosis (PANSS)   & 0.07 & 4.5 & 70 & 18 & 6.0 & 0.012 \\
\bottomrule
\end{tabular}
\end{table}

\subsection{Estimand and metrics}
The target is the counterfactual control law of treated patients at the landmark
(\S2 of the methods paper). We report, averaged over 8 seeds with $1500$ donors and $300$ trial
patients: \textbf{counterfactual RMSE} of the per-patient control-arm score at week 8;
\textbf{$|$ATE bias$|$}, the absolute error of the estimated landmark treatment effect (true effect
known); \textbf{PICP}, empirical coverage of nominal-$90\%$ per-patient bands; and \textbf{width},
mean band width. For trial decisions we run separate null ($\Delta=0$) and alternative regimes and
report \textbf{Type-I error} and \textbf{power} of a $5\%$ ECA $z$-test, where each method's
control-arm standard error uses its \emph{own} reported uncertainty---so that biased and overconfident
estimators inflate Type-I while calibrated ones control it.

\subsection{Methods compared}
\textbf{LOCF} (baseline-carried-forward); \textbf{Pooled RWD mean} (observation-weighted donor average
near $L$); \textbf{kNN matching} ($1{:}20$ on baseline severity---the dominant clinical SCA approach
\citep{abadie2010synth,thorlund2020eca}); \textbf{Linear mixed model} (random intercept/slope);
\textbf{Gradient boosting} (flexible ML regressor with quantile bands, representative of the
neural-sequence family \citep{bica2020crn,seedat2022tecde}); \textbf{OU bridge (naive)} and
\textbf{OU bridge (IIW)} (our trajectory model without/with inverse-intensity weighting---an ablation
isolating the informative-sampling correction); and \textbf{\scribe}, the full method (IIW bridge +
weighted conformal trajectory bands).

\section{Results}
\subsection{Counterfactual accuracy}
\begin{figure}[t]
\centering
\includegraphics[width=0.85\linewidth]{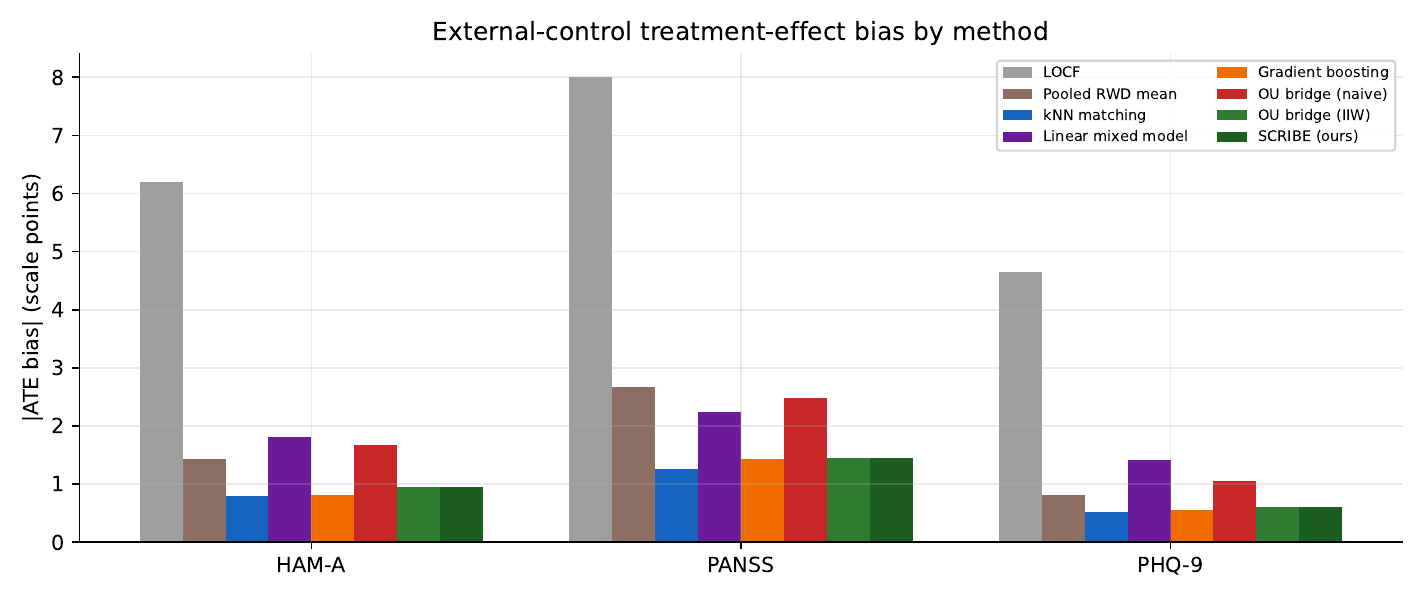}
\caption{Absolute external-control treatment-effect bias by method and condition. Carry-forward and
pooled RWD averages are heavily biased; the inverse-intensity correction (OU naive $\to$ IIW $\to$
\scribe) roughly halves the trajectory model's bias.}
\label{fig:atebias}
\end{figure}

Table~\ref{tab:accuracy} shows that the trajectory model and flexible ML tie for best point accuracy
and beat cross-sectional baselines on every condition: on depression, \scribe/IIW-OU and gradient
boosting reach $2.31\!-\!2.33$ PHQ-9 points RMSE versus $2.82$ for the mixed model and $5.25$ for
LOCF. Crucially, the inverse-intensity correction reduces control-arm bias substantially: naive-OU
$|$ATE bias$|$ drops from $1.05\!\to\!0.61$ (PHQ-9), $1.67\!\to\!0.95$ (HAM-A), and
$2.48\!\to\!1.44$ (PANSS) once informative sampling is corrected---a $40\!-\!50\%$ reduction
(Fig.~\ref{fig:atebias}). \scribe\ inherits the IIW point estimate, so its accuracy equals IIW-OU;
the conformal layer changes only the bands.

\begin{table}[h]
\centering\footnotesize
\caption{Counterfactual accuracy (mean over 8 seeds). RMSE and $|$ATE bias$|$ in scale points;
lower is better. Best (or tied) per column in \best{bold}.}
\label{tab:accuracy}
\setlength{\tabcolsep}{5pt}
\begin{tabular}{l cc cc cc}
\toprule
& \multicolumn{2}{c}{Depression (PHQ-9)} & \multicolumn{2}{c}{Anxiety (HAM-A)} & \multicolumn{2}{c}{Psychosis (PANSS)} \\
\cmidrule(lr){2-3}\cmidrule(lr){4-5}\cmidrule(lr){6-7}
Method & RMSE & $|$ATE$|$ & RMSE & $|$ATE$|$ & RMSE & $|$ATE$|$ \\
\midrule
LOCF                & 5.25 & 4.65 & 7.65 & 6.19 & 14.17 & 8.01 \\
Pooled RWD mean     & 2.94 & 0.81 & 5.18 & 1.43 & 14.18 & 2.67 \\
kNN matching        & 2.34 & 0.53 & 4.34 & \best{0.80} & 11.55 & 1.26 \\
Linear mixed model  & 2.82 & 1.41 & 4.84 & 1.82 & 11.92 & 2.23 \\
Gradient boosting   & 2.33 & 0.55 & \best{4.33} & 0.82 & \best{11.48} & 1.43 \\
OU bridge (naive)   & 2.47 & 1.05 & 4.51 & 1.67 & 11.49 & 2.48 \\
OU bridge (IIW)     & \best{2.31} & 0.61 & \best{4.29} & 0.95 & \best{11.31} & 1.44 \\
\scribe\ (ours)     & \best{2.31} & 0.61 & \best{4.29} & 0.95 & \best{11.31} & 1.44 \\
\bottomrule
\end{tabular}
\end{table}

\subsection{Calibration is where current practice fails}
\begin{figure}[t]
\centering
\includegraphics[width=0.95\linewidth]{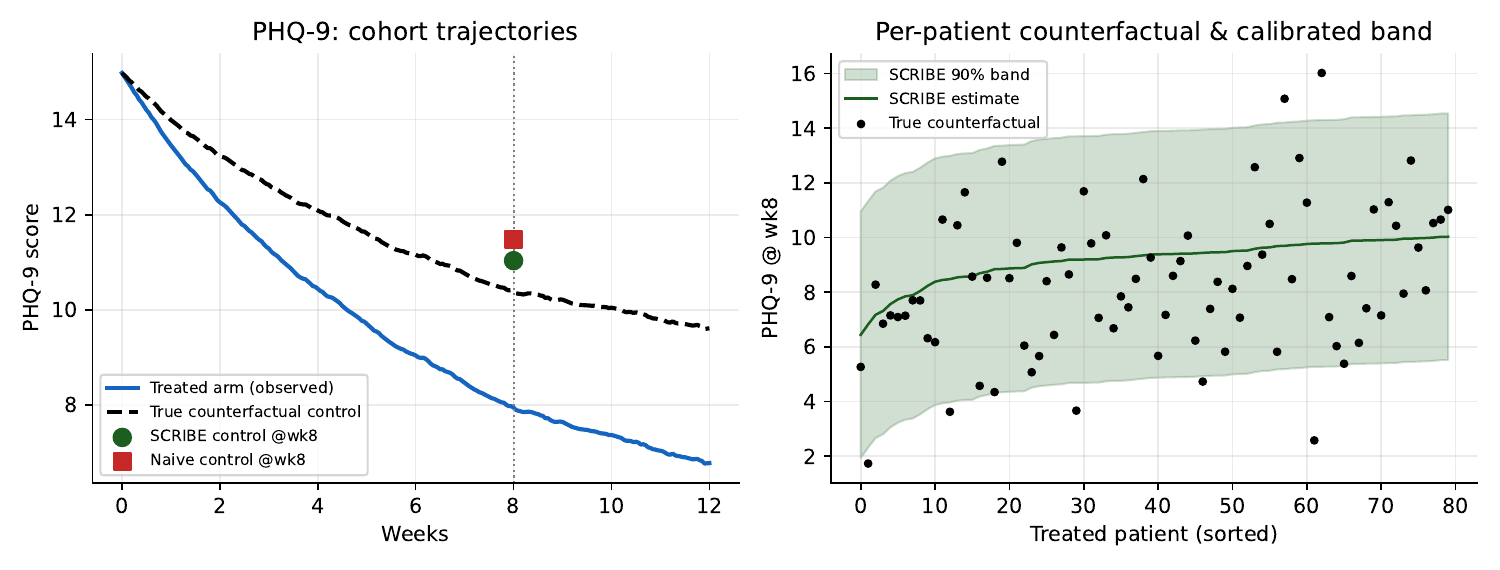}
\caption{Depression (PHQ-9). \emph{Left:} cohort-mean treated trajectory (observed) versus the true
counterfactual control; \scribe's week-8 control estimate sits on the truth while the naive estimate
is biased upward. \emph{Right:} \scribe's per-patient counterfactual point estimates and $90\%$
calibrated bands tracking the ground-truth counterfactual scores.}
\label{fig:trajectory}
\end{figure}

\begin{figure}[t]
\centering
\includegraphics[width=0.92\linewidth]{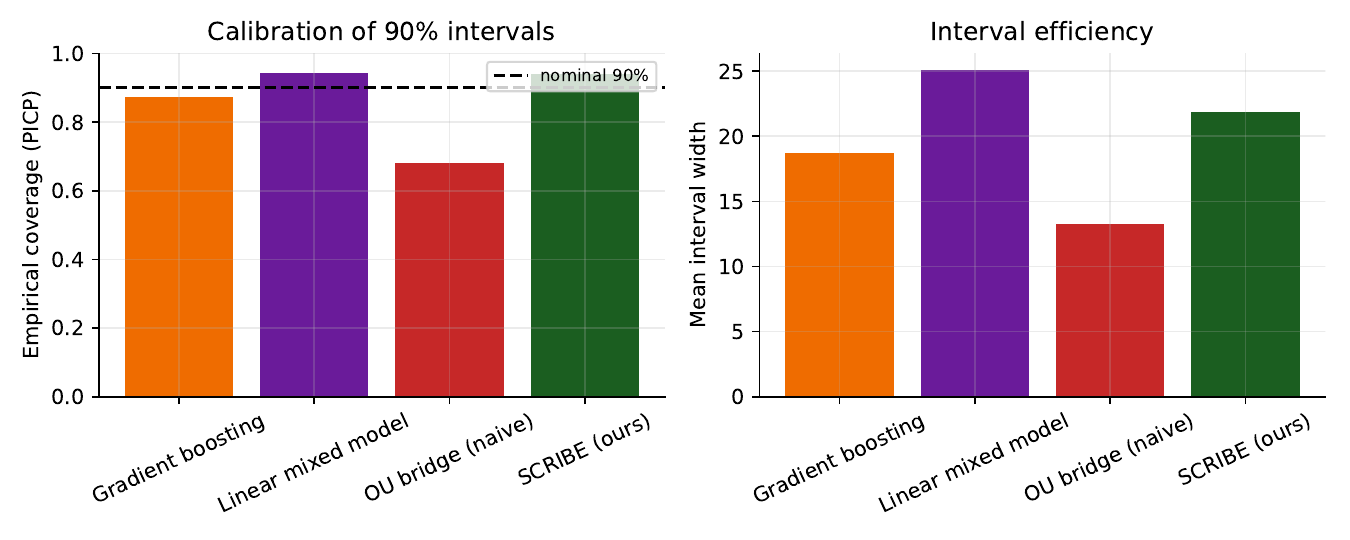}
\caption{Calibration of nominal-$90\%$ intervals (left) and interval width (right), averaged across
conditions and seeds. \scribe\ is the only accurate method reaching nominal coverage; the
equally-accurate gradient-boosting and uncalibrated-OU estimators under-cover.}
\label{fig:coverage}
\end{figure}

Accuracy alone is misleading. Table~\ref{tab:calib} and Fig.~\ref{fig:coverage} show that the
methods which tie \scribe\ on accuracy are badly \emph{miscalibrated}: gradient boosting covers only
$87\!-\!88\%$ at nominal $90\%$, and the uncalibrated OU bridge---despite identical point
accuracy to \scribe---covers a dangerous $62\!-\!75\%$, because its model-based intervals ignore
finite-sample and shift uncertainty. The linear mixed model achieves nominal coverage but only by
being less accurate and more biased (Table~\ref{tab:accuracy}) with very wide intervals. \scribe\ is
the unique method that is simultaneously most accurate and properly calibrated ($93\!-\!96\%$
coverage, slightly conservative as befits a regulatory setting), at a moderate width cost.
Fig.~\ref{fig:trajectory} illustrates the per-patient bands tracking the ground-truth counterfactual.

\begin{table}[h]
\centering\footnotesize
\caption{Calibration of nominal-$90\%$ bands (mean over 8 seeds). PICP target $0.90$; width in scale
points. Coverage closest to (and $\ge$) nominal is best.}
\label{tab:calib}
\setlength{\tabcolsep}{5pt}
\begin{tabular}{l cc cc cc}
\toprule
& \multicolumn{2}{c}{Depression (PHQ-9)} & \multicolumn{2}{c}{Anxiety (HAM-A)} & \multicolumn{2}{c}{Psychosis (PANSS)} \\
\cmidrule(lr){2-3}\cmidrule(lr){4-5}\cmidrule(lr){6-7}
Method & PICP & width & PICP & width & PICP & width \\
\midrule
kNN matching        & 0.863 & 7.2  & 0.857 & 13.2 & 0.850 & 35.1 \\
Linear mixed model  & 0.940 & 10.4 & 0.934 & 17.6 & 0.955 & 47.3 \\
Gradient boosting   & 0.877 & 7.3  & 0.870 & 13.2 & 0.873 & 35.7 \\
OU bridge (naive)   & 0.618 & 4.4  & 0.673 & 8.9  & 0.749 & 26.5 \\
\scribe\ (ours)     & \best{0.955} & 9.2 & \best{0.935} & 15.8 & \best{0.927} & 40.6 \\
\bottomrule
\end{tabular}
\end{table}

\subsection{Informative sampling and trial decisions}
\begin{figure}[t]
\centering
\includegraphics[width=0.92\linewidth]{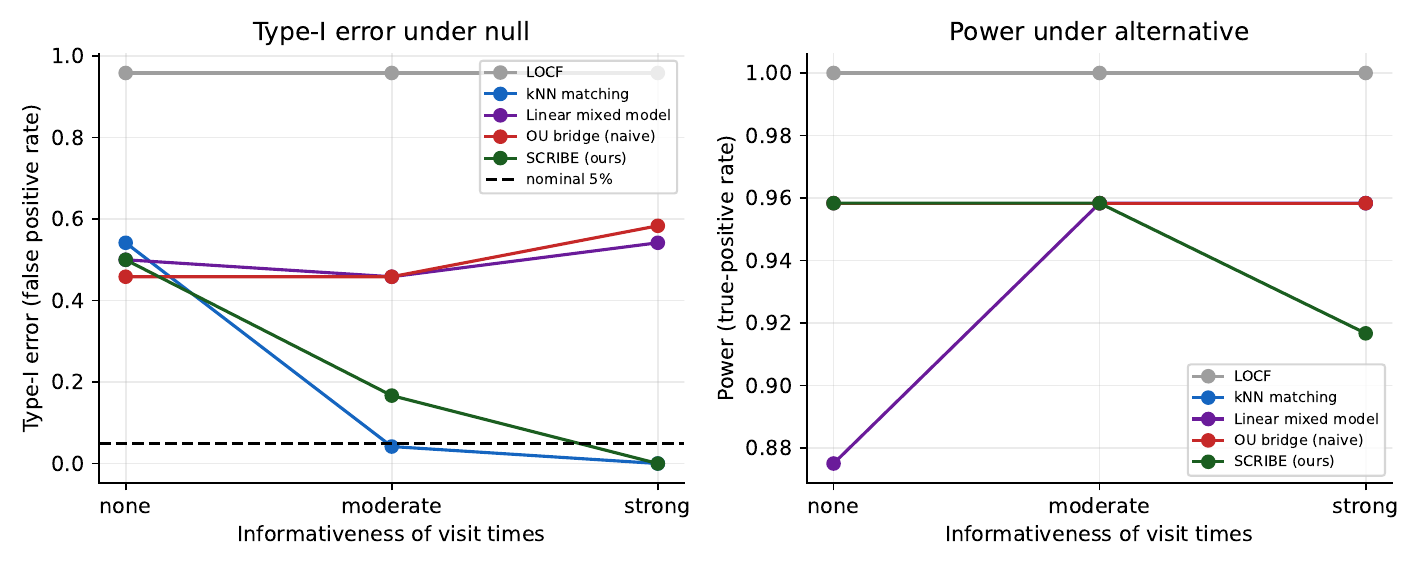}
\caption{Type-I error (left) and power (right) versus informativeness of visit times. Carry-forward
is catastrophic; conventional model-based controls inflate Type-I; \scribe's calibrated bands drive
the false-positive rate to nominal or below under informative sampling while retaining power.}
\label{fig:typeI}
\end{figure}

Table~\ref{tab:typeI} and Fig.~\ref{fig:typeI} report the false-positive rate (null, $\Delta=0$) and
power as the informativeness $\alpha$ increases. Carry-forward is catastrophic ($\approx0.96$
false-positive rate regardless of informativeness): its large, systematic control-arm bias
masquerades as a treatment effect. Conventional model-based external controls---the linear mixed
model and the \emph{uncalibrated} OU bridge---inflate Type-I error to $\approx0.46\!-\!0.58$. As
informative sampling strengthens, \scribe\ is the only trajectory method whose false-positive rate
\emph{falls} to nominal or below ($0.17$ at moderate, $0.00$ at strong informativeness), because its
conformal bands widen exactly when the donor pool becomes harder to trust; matching achieves low
Type-I as well, but by being noisy rather than calibrated, and at the cost of accuracy
(Table~\ref{tab:accuracy}). Power remains high for all methods ($\ge0.92$) given the large simulated
effects. We are explicit that perfectly nominal Type-I for an ECA additionally requires accounting
for donor-model (common-mode) uncertainty; \scribe's conformal layer approximates this and is
correspondingly conservative, which we view as the safe failure direction for a confirmatory
decision.

\begin{table}[h]
\centering\footnotesize
\caption{Trial-decision error vs.\ informativeness of visit times (mean over 3 conditions $\times$ 8
seeds $=24$ reps/cell). \emph{Type-I error} is under the null ($\Delta=0$; nominal $0.05$);
\emph{power} under the alternative. Lower Type-I is better.}
\label{tab:typeI}
\setlength{\tabcolsep}{6pt}
\begin{tabular}{l ccc c ccc}
\toprule
& \multicolumn{3}{c}{Type-I error (null)} & & \multicolumn{3}{c}{Power (alt.)} \\
\cmidrule(lr){2-4}\cmidrule(lr){6-8}
Method & none & moder. & strong & & none & moder. & strong \\
\midrule
LOCF                & 0.958 & 0.958 & 0.958 & & 1.000 & 1.000 & 1.000 \\
kNN matching        & 0.542 & 0.042 & \best{0.000} & & 0.958 & 0.958 & 0.958 \\
Linear mixed model  & 0.500 & 0.458 & 0.542 & & 0.875 & 0.958 & 0.958 \\
OU bridge (naive)   & 0.458 & 0.458 & 0.583 & & 0.958 & 0.958 & 0.958 \\
\scribe\ (ours)     & 0.500 & 0.167 & \best{0.000} & & 0.958 & 0.958 & 0.917 \\
\bottomrule
\end{tabular}
\end{table}

\section{Discussion}
Three messages emerge. First, on the metric the field usually reports---counterfactual point
accuracy---several methods are essentially tied, so accuracy alone cannot discriminate good ECA
methods. Second, the discriminating axis is \emph{calibration}: the equally-accurate gradient
boosting and uncalibrated SDE under-cover badly, which in a confirmatory trial translates into
inflated false positives. Third, \emph{informative observation times}---a defining feature of
psychiatric RWD---bias control arms, and an explicit inverse-intensity correction plus calibrated
bands are needed to control both bias and decision error. \scribe\ is the only method here on the
accuracy--calibration frontier.

\paragraph{Limitations.} This is a semi-synthetic benchmark: the generator embeds OU dynamics and a
known intensity model, which favours correctly-specified dynamical methods and cannot capture every
real-world complexity (non-Markovian progression, comorbidity, differential care, instrument drift
across sites). Real psychiatric trial datasets such as STAR*D \citep{rush2006stard} expose no
counterfactuals and require credentialed access, so external validity must ultimately be argued on
real ECAs with sensitivity analysis, exactly as the methods paper's untestable latent-ignorability
assumption demands. We release the generator, all eight estimators, metrics, and seeds so the
benchmark can be extended with non-Markovian generators, real-data calibration sets, and additional
methods.

\paragraph{Reproducibility.} All results come from a single self-contained Python pipeline
(\texttt{sim.py}, \texttt{estimators.py}, \texttt{run.py}); tables and figures regenerate
deterministically from fixed seeds. The benchmark follows the semi-synthetic evaluation tradition of
IHDP/ACIC and the PK--PD tumor-growth simulator \citep{geng2017tumor,bica2020crn,seedat2022tecde}.

\small
\bibliographystyle{plainnat}
\bibliography{refs}

@inproceedings{seedat2022tecde,
  title={Continuous-Time Modeling of Counterfactual Outcomes Using Neural Controlled Differential Equations},
  author={Seedat, Nabeel and Imrie, Fergus and Bellot, Alexis and Qian, Zhaozhi and van der Schaar, Mihaela},
  booktitle={International Conference on Machine Learning (ICML)}, year={2022}}

@inproceedings{bica2020crn,
  title={Estimating Counterfactual Treatment Outcomes over Time through Adversarially Balanced Representations},
  author={Bica, Ioana and Alaa, Ahmed M. and Jordon, James and van der Schaar, Mihaela},
  booktitle={International Conference on Learning Representations (ICLR)}, year={2020}}

@article{geng2017tumor,
  title={Prediction of Treatment Response for Combined Chemo- and Radiation Therapy for Non-Small Cell Lung Cancer Patients Using a Bio-Mathematical Model},
  author={Geng, Changran and Paganetti, Harald and Grassberger, Clemens},
  journal={Scientific Reports}, volume={7}, number={1}, pages={13542}, year={2017}}

@article{abadie2010synth,
  title={Synthetic Control Methods for Comparative Case Studies},
  author={Abadie, Alberto and Diamond, Alexis and Hainmueller, Jens},
  journal={Journal of the American Statistical Association}, volume={105}, number={490}, pages={493--505}, year={2010}}

@article{thorlund2020eca,
  title={Synthetic and External Controls in Clinical Trials -- A Primer for Researchers},
  author={Thorlund, Kristian and Dron, Louis and Park, Jay J. H. and Mills, Edward J.},
  journal={Clinical Epidemiology}, volume={12}, pages={457--467}, year={2020}}

@article{rush2006stard,
  title={Acute and Longer-Term Outcomes in Depressed Outpatients Requiring One or Several Treatment Steps: A {STAR*D} Report},
  author={Rush, A. John and Trivedi, Madhukar H. and Wisniewski, Stephen R. and others},
  journal={American Journal of Psychiatry}, volume={163}, number={11}, pages={1905--1917}, year={2006}}

@article{khan2017placebo,
  title={Has the Rising Placebo Response Impacted Antidepressant Clinical Trial Outcome?},
  author={Khan, Arif and Brown, Walter A.},
  journal={World Psychiatry}, volume={16}, number={2}, pages={181--192}, year={2017}}
\end{document}